# DETERMINATION DU CENTRE DE RAIDEUR POUR LES MACHINES-OUTILS - APPLICATION AU TOURNAGE

**Claudiu Florinel BISU**
LMS, Université POLITEHNICA de Bucarest,
Splaiul Independentei 313 Bucarest – Roumanie, 0040724016295, cfbisu@gmail.com.

**Jean Yves K'NEVEZ**
LMP, Université Bordeaux1, 351 Cours de la Libération, 33405 Talence Cedex - France,
tél : +(33) 05 40 00 89 28, fax : +(33) 5 40 00 69 64, jean-yves.knevez@u-bordeaux1.fr

**Philippe Darnis**
$LGM^2B$, IUT Bordeaux 1, Domaine Universitaire33405 Talence Cedex - France,
tél. : +(33) 5 40 00 79 76, fax : +(33) 5 56 84 58 43, philippe.darnis@u-bordeaux1.fr

**Olivier CAHUC**
LMP, Université Bordeaux1, 351 Cours de la Libération, 33405 Talence Cedex - France,
tél : +(33) 5 40 00 87 89, fax : +(33) 5 40 00 69 64, olivier.cahuc@u-bordeaux1.fr

**Raynald LAHEURTE**
$LGM^2B$, IUT Bordeaux 1, Domaine Universitaire33405 Talence Cedex - France,
tél. : +(33) 5 40 00 79 76, fax : +(33) 5 56 84 58 43, philippe.darnis@u-bordeaux1.fr

**Alain GERARD**
LMP, Université Bordeaux1, 351 Cours de la Libération, 33405 Talence Cedex - France,
tél : +(33) 05 40 00 89 28, fax : +(33) 5 40 00 62 23, alain.gerard@u-bordeaux1.fr

*Résumé :* La détermination du centre de raideur d'une machine-outils est un élément important pour prédire le comportement dynamique de celle-ci. Ce centre de raideur peut être déterminé par une méthode expérimentale de caractérisation statique. Pour une opération de tournage, nous présentons un protocole expérimental pour déterminer la matrice de raideur, puis par inversion la matrice de souplesse associée. Cette démarche permet d'établir les directions privilégiées du mouvement de l'outil par rapport à la pièce et de localiser le centre de raideur. Au travers de cette démarche, il est observé que, lors d'une opération de tournage, le mouvement de la pointe de l'outil s'effectue suivant une ellipse située dans un plan incliné par rapport aux axes de la machine-outils. Ces mouvements sont mis en évidence avec la détection de vibrations auto-entretenues. Cette méthode pourra être exploitée et transposée dans le cas de l'UGV. Elle permettra de prédire divers aspects de l'usinage liés au comportement de l'outil dans le processus de coupe.

**Mots clés :** *Centre de raideur, Matrice de raideur, Matrice des souplesses*

*Abstract*: The determination of the center of stiffness of a machine tool is a major element for knowing the dynamic behavior of this one well. This center of stiffness can result starting from a static operation of characterization from a machine tool. For an operation of turning, we present an experimental protocol to determine the matrix of stiffness, then by inversion the matrix of associated flexibility. This step makes it possible to establish the privileged directions of the movement associated with the maximum and minimal stiffnesses as well as the center with stiffness. Through this step, it is observed that, at the time of an operation of turning, the movement of the point of the tool is carried out according to an ellipse located in a tilted average plan compared to the machine spindles. This movement is in connection with the appearance of the self-excited vibrations. This analysis could be exploited and transposed in the case of the HSM. It will make it possible to predict various aspects of machining related to the behavior of the tool in the process of cut.

**1 Introduction**

De nos jours, les machines-outils sont très rigides, et possèdent de moins en moins de défauts géométriques. Les problèmes dynamiques (vibrations) sont fortement liés à la coupe. Ces vibrations sont générées et auto-entretenues par le processus de coupe.

Dans l'objectif de développer un modèle dynamique tridimensionnel de la coupe la caractérisation du système usinant est nécessaire. L' étude dynamique expérimentale présentée est divisée en deux parties ; le système usinant, et le processus de coupe. La dissociation de ces deux parties (système usinant / processus de coupe), est nécessaire pour modéliser les phénomènes vibratoires. Cette démarche permet d'établir les directions privilégiées du mouvement associées aux raideurs maximales et minimales ainsi que le centre de raideur.

La structure élastique de la machine-outils, le système pièce/outil/machine (**POM**), est un système à plusieurs degrés de liberté. Il possède un grand nombre de modes propres de vibration. Le comportement vibratoire de chaque sous ensembles de la structure, est caractérisé par sa fréquence propre fonction de sa raideur $[K]$, de sa masse $[M]$ et de son coefficient d'amortissement $[C]$. La structure élastique du système **POM** est caractérisée par un nombre élevé de couples cinématiques et d'ensembles avec serrage réduit [1],[2]. Pour identifier le comportement vibratoire du système **POM**, celui-ci est divisé en deux blocs, le bloc outil (**BO**) et le bloc pièce (**BP**), figure 1, [3].

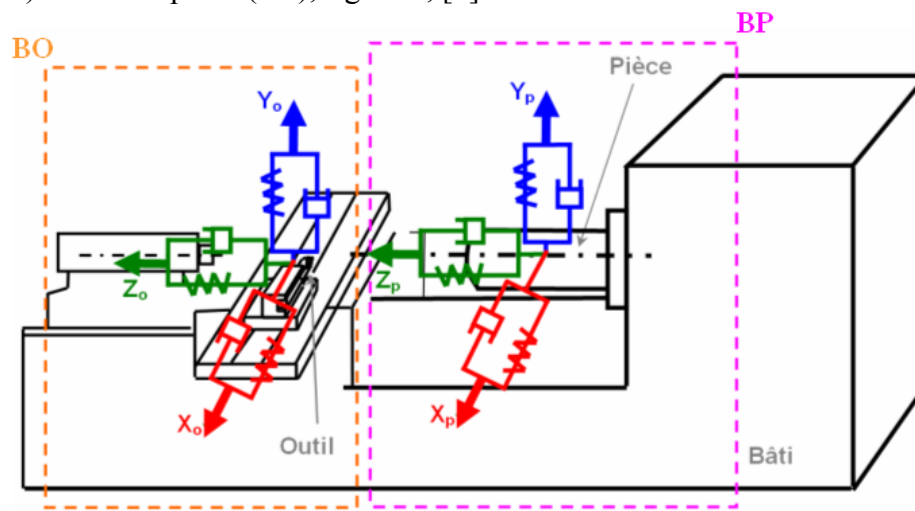

*Fig. 1 Présentation du dispositif expérimental divisé en : **BO** et **BP**.*

**2 Protocole expérimentale**

La caractérisation statique du système usinant est effectuée en deux temps : la détermination du comportement élastique de celui-ci et l'identification des paramètres qui influencent l'apparition des vibrations lors de la coupe. Un protocole expérimental est mis en place pour munir le modèle les raideurs nécessaires à la corrélation des efforts et des déplacements de l'outil lors de la coupe. Pour étudier ces phénomènes dynamiques, nous disposons d'un tour conventionnel possédant une rigidité élevée. Les essais sont effectués pour chaque direction de mesure, x, y et z (directions liées à la machine). Les déplacements et les rotations sont mesurés lorsque la force est appliquée. La force imposée est contrôlée à l'aide d'un capteur d'effort unidirectionnel et d'un afficheur électronique, tandis que le torseur d'action complet et le torseur des petits déplacements sont enregistrés à partir d'un dynamomètre à 6 composantes. Le torseur des petits déplacements est déterminé par six capteurs de micro-déplacements constituant la chaîne de mesure associée implémentée sous le logiciel Labview. Dans la figure 2 sont présentés le dispositif expérimental, et les directions des chargements retenues [4].

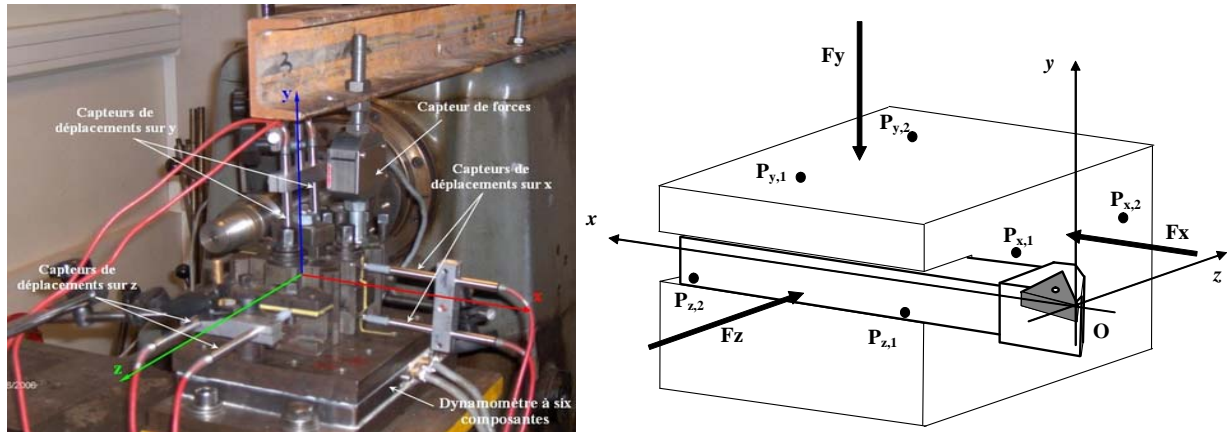

*Fig.2: Chargement sur x, y, z*

Les 6 composantes du torseur de déplacements sont calculées à partir de 6 mesures. Pour exploiter au mieux les mesures, les courbes de déplacements sont tracées en fonction de la force appliquée en un point particulier, correspondant à chaque direction de charge. Une droite des moindres carrés est ajustée pour déterminer les valeurs des composantes des déplacements pour une force donnée. Pour 6 cas de chargement nous déterminons 6 torseurs de petits déplacements « lissés ». Or le système usinant est constitué d'une multitude d'assemblages et de jeux. Des comportements non linéaires apparaissent, en particulier de l'hystérésis est détectée entre les courbes de charge-décharge. Cette hystérésis est due aux forces de frottement existantes dans chaque assemblage.

De l'exploitation de la méthode des moindres carrées pour les deux matrices et de l'évaluation de la courbe de déplacements, nous déduisons le torseur des actions mécaniques et le torseur des petits déplacements. Par ailleurs nous avons évalué les incertitudes de mesures pour le dynamomètre à ±5N et pour les capteurs de déplacements à ±2µm. La construction effective de ces deux torseurs est réalisée lors des chargements en six points de coordonnées connues suivant des directions connues. Dans ces conditions, nous relevons les déplacements des six points, et déduisons le torseur des petits déplacements.

## 3 Comportement dynamique du BO

Dans la majorité des processus dynamiques, la liaison entre le processus et le système élastique se manifeste par l'action de la force générée lors de l'enlèvement de matière. Plus exactement la déformation plastique du matériau enlevé sous forme de copeaux génère les forces et les moments de coupe. Cette déformation est liée au processus qui a lieu sur les surfaces de l'outil en contact avec le copeau et la pièce, et donc avec le système usinant **POM**.

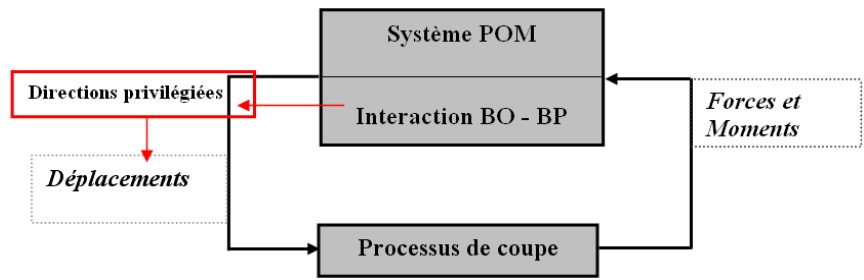

*Fig.3 : Diagramme de comportement dynamique du **POM**.*

Le comportement statique du **BO** et du **BP**, est identifiable sur le comportement dynamique par l'interaction de **BO – BP**. Des directions privilégiées du système élastique influencent les directions principales du mouvement de l'outil lors de la coupe. Cette étude se concentre sur la caractérisation en statique du système usinant dans le but d'analyser les phénomènes complexes et non linéaires qui influencent le caractère dynamique de l'outil pendant la coupe.
Le principe est de trouver un outil qui permet de donner une réponse sur le comportement élastique du système, et de trouver les directions privilégiées des déplacements 3D.

### 3.1 Détermination du centre de raideur du bloc outil

Des auteurs [5], [6] mettent en évidence que la structure dynamique d'une machine-outils, ou d'une partie de cet ensemble, présente des directions, qui, en fonction de la répartition des masses et des configurations géométriques, a un comportement soit très raide, soit très élastique. Connaissant la disposition des éléments de mesures et le protocole qui fournit des informations 3D, nous déterminons le centre de rotation. Pour ce faire, nous considérons l'outil comme faisant partie intégrante de la partie **BO**. Par la suite les points de chargement (figure 2) sont choisis de manière à avoir une réponse propre du système à la force imposée.

Le comportement élastique lors de la coupe doit être étudié pour tout l'ensemble **BO** et pas seulement pour l'outil. Le système élastique du **BO** est schématisé par un système possédant un centre de rotation, ou centre de raideur $(CR_{BO})$ du système **BO** [7].

La méthode employée pour chercher le $CR_{BO}$ consiste à trouver les points d'intersection suivant chaque direction lors des chargements appliqués. Comme chaque point est situé dans un plan, trois plans sont définis. Le point d'intersection des perpendiculaires à chaque plan définit le centre de raideur $CR_{BO}$. Les essais sont réalisés dans chaque direction à partir d'un chargement $(F)$. Deux points sont choisis sur chaque direction sur la direction du **BO**. La pointe de l'outil est le point de O(0, 0, 0) figure 5. En chaque point de chargement est défini un vecteur des déplacements $D_{ij}$ à partir des coordonnées de chaque point repéré par rapport à la pointe de l'outil. Par exemple, pour le point de chargement $X_2 = (X_{2x}, X_{2y}, X_{2z})$, nous avons le vecteur $\vec{D}_{2x} = (\varepsilon_{2xx}, \varepsilon_{2xy}, \varepsilon_{2xz})$. Ensuite pour chaque direction, nous déterminons les coordonnées du point et le vecteur déplacement correspondant à cette direction, puis nous cherchons le point d'intersection pour chaque direction x, y et z:

$$A(e_x) = X_2 + e_x \vec{D}_{2x}, \quad a(e_x) = \frac{\left| A((e_x) - X_3) \cdot \vec{D}_{3x} \right|}{\left\| \vec{D}_{3x} \right\|} \tag{1}$$

$$B(e_y) = Y_2 + e_y \vec{D}_{2y}, \quad b(e_y) = \frac{\left|(B(e_y) - Y_3) \cdot \vec{D}_{3y}\right|}{\left\|\vec{D}_{3y}\right\|} \quad (2)$$

$$C(e_z) = Z_2 + e_z \vec{D}_{2z}, \quad c(e_z) = \frac{\left|(C(e_z) - Z_3) \cdot \vec{D}_{3z}\right|}{\left\|\vec{D}_{3z}\right\|} \quad (3)$$

Par application de la méthode des moindres carrés, nous trouvons le point d'intersection correspondants à chaque direction de chargement, notées ; $J_x(e_x)$ sur x, $J_y(e_y)$ sur y et $J_z(e_z)$ sur z (figure 4).

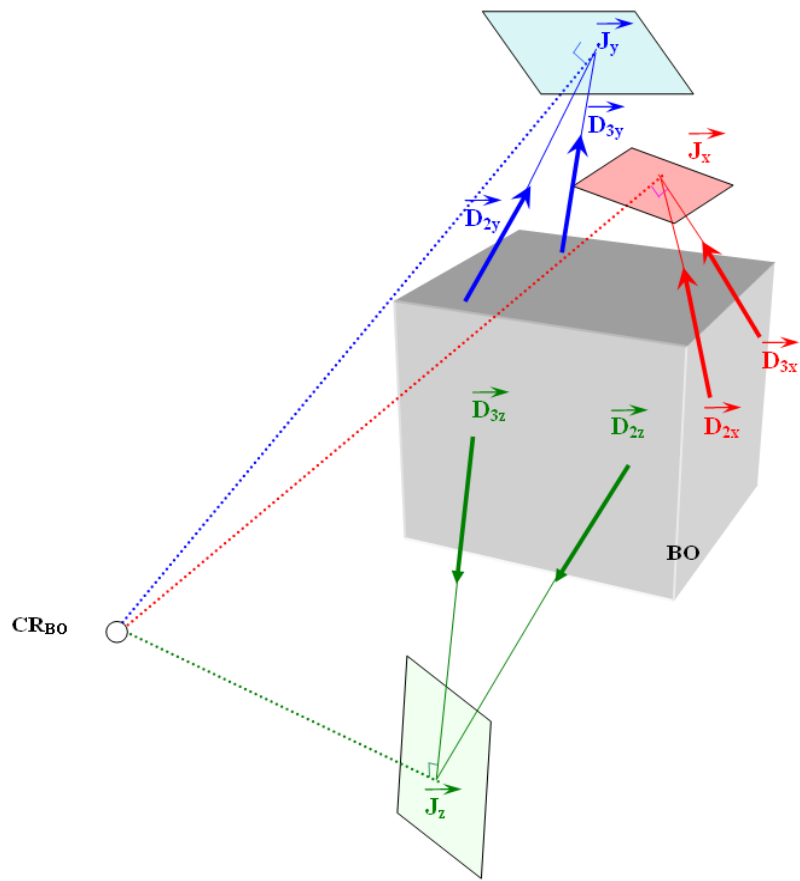

*Fig.4 : Détermination du centre de raideur $CR_{BO}$.*

Si la coupe est orthogonale, le point d'intersection de deux directions lors du chargement représente le centre de raideur [6]. Mais comme nous travaillons en coupe tridimensionnelle et que le comportement du système est aussi 3D, le centre de raideur doit être le point de concours de toutes les directions. Par cette méthode, nous déterminons la droite qui passe par la pointe de l'outil et par chaque point d'intersection suivant les trois directions, et ensuite les plans formés par ces points et ces droites. Pour déterminer les coordonnées du centre de raideur, nous écrivons le système d'équations à l'aide des perpendiculaires à ces plans au point d'intersection correspondant à chaque direction. Le centre de raideur correspond au point d'intersection de ces perpendiculaires.

$$\begin{cases} \left(CR_{BO}(x,y,z)-J_x(e_x)\right)\cdot\left(J_x(e_x)-O\right)=0 \\ \left(CR_{BO}(x,y,z)-J_y(e_y)\right)\cdot\left(J_y(e_y)-O\right)=0 \\ \left(CR_{BO}(x,y,z)-J_z(e_z)\right)\cdot\left(J_z(e_z)-O\right)=0 \end{cases} \quad (4)$$

### 3.2 Localisation de centre de raideur

La résolution de ce système linéaire de 3 équations à 3 inconnues permet de définir le point d'intersection de ces perpendiculaires. Nous obtenons ainsi le centre de raideur ou de rotation du **BO**, noté $CR_{BO}$. Les valeurs de ce calcul sont présentées dans le tableau 1. Vu les valeurs de l'intersection suivant chaque axe, nous attribuons les décalages du point d'intersection aux erreurs de mesures et aux différentes hystérésis mis en jeu. Néanmoins, les résultats obtenus par cette démarche expérimentale sont cohérents avec ceux de la littérature [6].

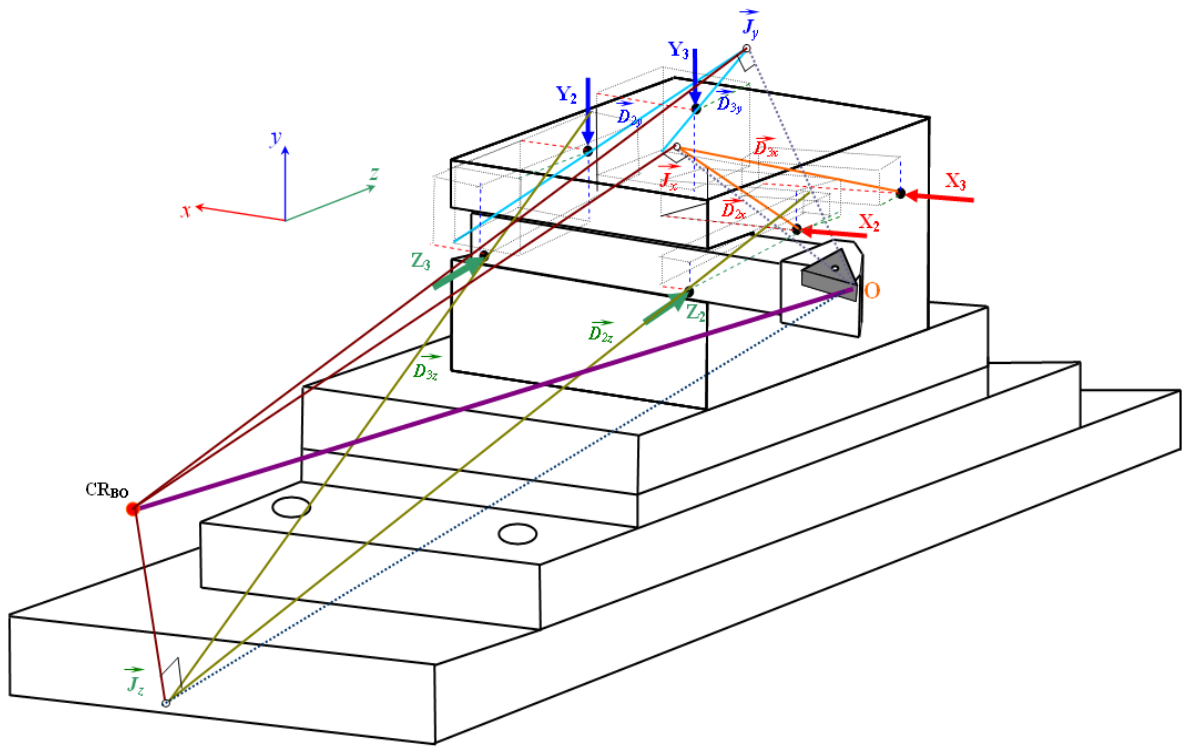

*Fig.5: Localisation du centre de raideur dans la partie **BO**.*

Notons bien que ce type d'analyse expérimentale est réalisé dans des conditions tridimensionnelles, ce qui est nécessaire pour pouvoir expliciter par la suite le comportement dynamique de la coupe tridimensionnelle.

| Direction | Point de chargement | Coordonnées du point *(mm)* | Vecteur déplacement *(m)* | Décalage d'intersection *(m)* | Point $J$ intersection *(m)* | $CR_{BO}$ *(m)* |
|---|---|---|---|---|---|---|
| x | $X_2$ | 35 | $9.1 \cdot 10^{-5}$ | $1.8 \cdot 10^{-3}$ | 0.366 | 0.56 |
| | | -20 | $1.7 \cdot 10^{-5}$ | | 0.042 | |
| | | 52 | $3.4 \cdot 10^{-5}$ | | 0.178 | |
| | $X_3$ | 35 | $.8 \cdot 10^{-5}$ | | | |
| | | -20 | $1.5 \cdot 10^{-5}$ | | | |
| | | 117 | $1.5 \cdot 10^{-5}$ | | | |
| y | $Y_2$ | 116 | $2 \cdot 10^{-5}$ | $1.7 \cdot 10^{-3}$ | 0.086 | -0.58 |
| | | 15 | $-2 \cdot 10^{-5}$ | | 0.045 | |
| | | 56 | $-2 \cdot 10^{-5}$ | | 0.081 | |
| | $Y_3$ | 116 | $.2 \cdot 10^{-5}$ | | | |
| | | 15 | $-1.3 \cdot 10^{-5}$ | | | |
| | | 103 | $9.8 \cdot 10^{-6}$ | | | |
| z | $Z_2$ | 45 | $5.5 \cdot 10^{-6}$ | $8.8 \cdot 10^{-4}$ | 0.033 | -0.08 |
| | | -20 | $6.5 \cdot 10^{-5}$ | | -0.052 | |
| | | 17 | $5.5 \cdot 10^{-5}$ | | -0.257 | |
| | $Z_3$ | 130 | $3 \cdot 10^{-5}$ | | | |
| | | -20 | $1 \cdot 10^{-5}$ | | | |
| | | 6 | $8.7 \cdot 10^{-5}$ | | | |

*Tab.1 : Valeurs lors de la détermination du centre de raideur $CR_{BO}$.*

## 4 Interaction Bloc Outil/Bloc Pièce

Dans le but de connaître la direction principale de déformation du système usinant nous modélisons l'interaction élastique outil/pièce, **BO/BP** par un modèle de ressorts montés en parallèles (figure 6).

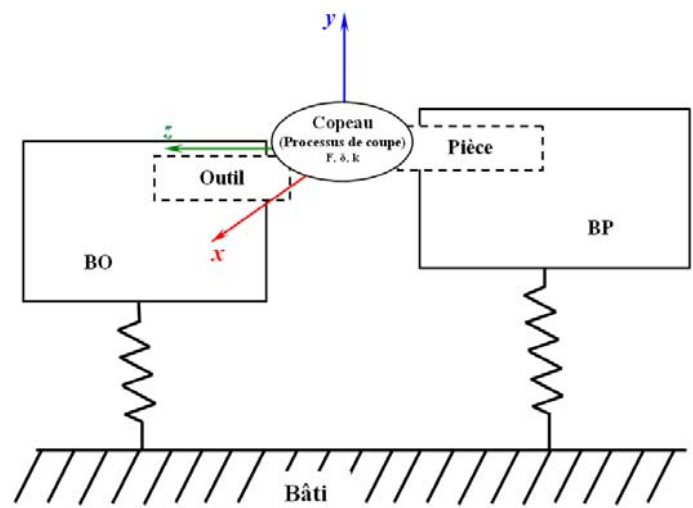

*Fig.6: Raideurs connectées en parallèle.*

Nous schématisons les raideurs statiques de l'ensemble à déterminer, le copeau étant le point commun qui ferme le système (figure 6). La direction principale de déformation lors de l'interaction de ces deux parties **BO** et **BP** est déterminée à l'aide de la théorie de Castigliano [8], figure 7. Lors d'expérimentation nous avons déterminé les matrices de raideur pour le **BO** et **BP** :

$$K_{BO} = \begin{bmatrix} 6.7 \cdot 10^6 & 8.7 \cdot 10^5 & -3.4 \cdot 10^6 \\ -7.5 \cdot 10^6 & -3.3 \cdot 10^5 & 1.7 \cdot 10^6 \\ -1.4 \cdot 10^6 & -1.7 \cdot 10^6 & 1.3 \cdot 10^7 \end{bmatrix} \quad (5)$$

Après diagonalisation nous avons :

$$K_{BO\_d} = \begin{bmatrix} 4.1 \cdot 10^5 & 4.6 \cdot 10^{-9} & -3.1 \cdot 10^{-9} \\ 0 & 6 \cdot 10^6 & 4.7 \cdot 10^{-10} \\ 4.7 \cdot 10^{-10} & -4.1 \cdot 10^{-10} & 1.3 \cdot 10^7 \end{bmatrix} \quad (6)$$

et en suite la matrice de raideur pour **BP** :

$$[K_{BP}] = \begin{bmatrix} 1,4 \cdot 10^7 & 0 & 0 \\ 0 & 2 \cdot 10^7 & 0 \\ 0 & 0 & 2.85 \cdot 10^8 \end{bmatrix} \quad (7)$$

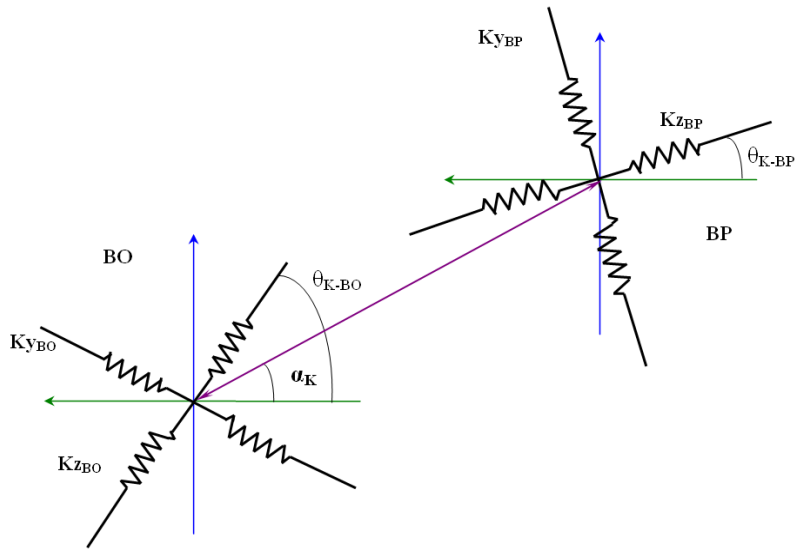

*Fig.7: Schéma de détermination d'angle α entre le système **BO** et **BP**.*

$$\alpha_K = arc\tan\left(\frac{a_1 \cdot \sin(2 \cdot \theta_{K-BO}) + a_2 \cdot \sin(2 \cdot \theta_{K-BP})}{a_1 \cdot \cos(2 \cdot \theta_{K-BO}) + a_2 \cdot \cos(2 \cdot \theta_{K-BP})}\right) \quad (8)$$

avec,

$$a_2 = \frac{1}{Ky_{BO}} - \frac{1}{Kz_{BO}}, \ a_1 = \frac{1}{Ky_{BP}} - \frac{1}{Kz_{BP}} \tag{9}$$

Pour la configuration du système $y,z$, avec $\theta_{K-BO} = 52°$ et $\theta_{K-BP} = 0°$ nous obtenons un angle $\alpha_K = 76°$, figure 7, tandis que dans la configuration $y,x$ avec $\theta_{K-BO} = 32°$ et $\theta_{K-BP} = 0°$ nous avons un angle $\alpha_K = 65°$.

**5 Corrélation déplacement de l'outil/centre de raideur**

A l'aide de l'analyse statique développée nous mettons en évidence le couplage entre les caractéristiques élastiques du système usinant et les vibrations générées par la coupe. Comme nous pouvons nous y attendre, l'apparition des vibrations auto-entretenues est fortement influencée par les valeurs des raideurs du système, leur rapport et leur direction. Lors de l'analyse des données accélérométriques pendant la coupe permet d'établir l'existence d'un plan des déplacements dans lequel la pointe de l'outil décrit une ellipse (figure 8).

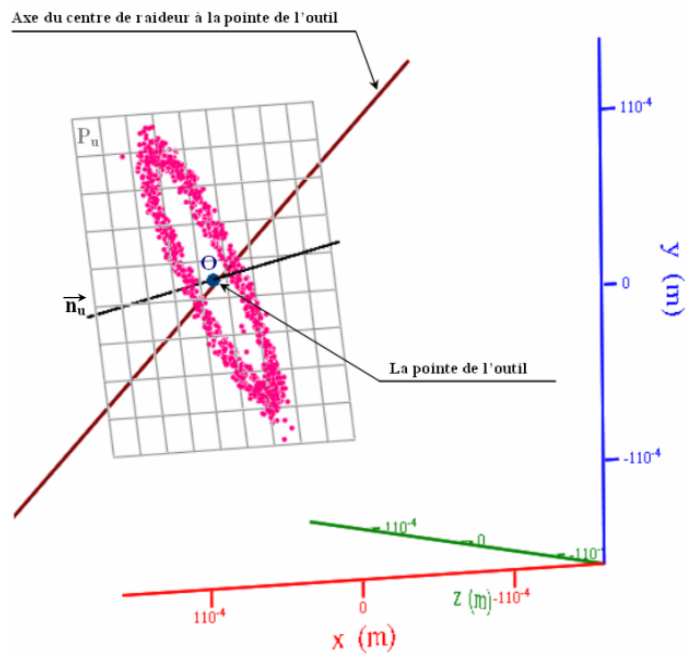

*Fig.8: Corrélation entre le centre de raideur et l'ellipse des déplacements.*

Dans la figure 8 nous traçons cet axe de raideur maximale qui est orthogonal à la normale au plan, et qui est aussi perpendiculaire au grand axe de l'ellipse. Ceci est conforme au fait que le plus grand déplacement ait lieu selon l'axe où la raideur est minimale. Cette corrélation valide notre protocole expérimental. La détermination de ce plan des déplacements et la caractérisation elliptique des mouvements dans ce plan, nous fournit des informations sur la configuration à adopter pour l'écriture du modèle et permet d'exprimer le système en fonction des axes de l'ellipse.

A partir de l'étude de l'angle de l'inclinaison de l'ellipse (figure 8), nous vérifions que les angles de la direction principale des déplacements sont comparables aux angles des ellipses des déplacements pour différentes avances de coupe *f* (tableau 2).

| Statique | | Dynamique | | |
|---|---|---|---|---|
| $α_K$ (yz) | $α_K$ (yx) | L'angle de l'ellipse (yz) | L'angle de l'ellipse (yx) | $f$ (mm/tr) |
| 76° | 65° | 79° | 65° | 0.05 |
| | | 79° | 67° | 0.0625 |
| | | 77° | 69° | 0.075 |
| | | 77° | 69° | 0.1 |

*Tab.2: Comparaison des angles de la direction principale des déplacements.*

## 4 Conclusions

Par rapport à d'autres auteurs qui utilisent uniquement les raideurs de l'outil, cette étude prend en compte, en plus, les raideurs issues du système de maintien de l'outil (**BO**). Des informations importantes sur les raideurs statiques sont obtenues, permettant notamment de déterminer les directions privilégiées des déplacements. Le centre de raideur et le centre de rotation ont été déterminés expérimentalement. La direction de déplacement minimum a également été définie à partir du modèle expérimental. Puis la partie **BP** est caractérisée. Ses raideurs sont déterminées et une matrice diagonale de déplacement est obtenue. L'analyse du comportement statique du système **BO - BP** est validée. Le protocole expérimental utilisé dans la détermination des raideurs statiques est aussi validé par les essais de coupe dans le domaine de la coupe vibratoire.

Cette analyse pourra être exploitée et transposée dans le cas de l'UGV. Elle permettra de prédire divers aspects de l'usinage liés au comportement de l'outil dans le processus de coupe sachant les directions privilégies des déplacements maximum (ou minimum) sont bien liées aux directions des raideurs minimales (ou maximales).

## 5 Références bibliographiques